\newif\ifpreprint

\preprintfalse 

\ifpreprint
\documentclass[aip,jcp,amsmath,amssymb,preprint]{revtex4-1}
\else
\documentclass[aip,jcp,amsmath,amssymb,reprint]{revtex4-1}
\fi

\usepackage[english]{babel}

\usepackage{graphicx}
\usepackage{dcolumn}
\usepackage{bm}
\usepackage{lineno}
\usepackage{amsfonts}
\usepackage{pbox}
\usepackage{braket}
\usepackage{multirow}
\usepackage{threeparttable}
\usepackage{float}
\usepackage{xspace}
\usepackage{color}
\usepackage{fancyvrb}
\usepackage{adjustbox}

\usepackage{booktabs}
\usepackage{tabularx}

\newcommand*{\abinitio}{{\it ab initio}\xspace}

\newcommand*{\kcal}{kcal mol$^{-1}$\xspace}
\newcommand*{\sunit}{$E_\mathrm{h}^{-2}$\xspace}
\newcommand*{\Eh}{$E_{\rm h}$\xspace}
\newcommand*{\Eht}{$E_{\rm h}^{-2}$\xspace}

\newcommand*{\PSI}{{\scshape Psi4}\xspace}
\newcommand*{\ambit}{{\scshape Ambit}\xspace}
\newcommand*{\nwchem}{{\scshape NWChem}\xspace}
\newcommand{\mref}[0]{\Phi}
\newcommand{\tens}[3]{{#1}_{#2}^{#3}}
\newcommand{\dfock}[1]{\varepsilon_{#1}}
\newcommand{\cop}[1]{\hat{a}^\dag_{#1}}
\newcommand{\aop}[1]{\hat{a}_{#1}}
\newcommand{\sqop}[2]{\hat{a}_{#2}^{#1}}

\newcommand{\kro}[2]{\delta_{#2}^{#1}}
\newcommand{\density}[2]{\gamma_{#2}^{#1}}
\newcommand{\cdensity}[2]{\eta_{#2}^{#1}}
\newcommand{\cumulant}[2]{\lambda_{#2}^{#1}}
\newcommand{\no}[1]{ \{ {#1} \}}

\newcommand{\methodname}[0]{{DSRG-MRPT2}\xspace}
\newcommand{\rv}[0]{\tilde{v}}

\begin{document}

\title{An integral-factorized implementation of the driven similarity renormalization group second-order multireference perturbation theory}

\author{Kevin P. Hannon}
\author{Chenyang Li}
\author{Francesco A. Evangelista}
\email{francesco.evangelista@emory.edu}
\affiliation{Department of Chemistry and Cherry Emerson Center for Scientific Computation, Emory University, Atlanta, GA, 30322}

\date{\today}

\begin{abstract}
We report an efficient implementation of a second-order multireference perturbation theory based on the driven similarity renormalization group (DSRG-MRPT2) [C.~Li and F.~A.~Evangelista, J.~Chem.~Theory~Comput.~\textbf{11}, 2097 (2015)].
Our implementation employs factorized two-electron integrals to avoid storage of large four-index intermediates.
It also exploits the block structure of the reference density matrices to reduce the computational cost  to that of second-order M{\o}ller--Plesset perturbation theory.
Our new DSRG-MRPT2 implementation is benchmarked on ten naphthyne isomers using basis sets up to quintuple-$\zeta$ quality.
We find that the singlet-triplet splittings ($\Delta_\text{ST}$) of the naphthyne isomers strongly depend on the equilibrium structures.
For a consistent set of geometries, the $\Delta_\text{ST}$ values predicted by the DSRG-MRPT2 are in good agreements with those computed by the reduced multireference coupled cluster theory with singles, doubles, and perturbative triples.
\end{abstract}

\maketitle

\section{Introduction}
Second-order M{\o}ller--Plesset perturbation theory (MP2) is perhaps one of the simplest approach to treat \textit{dynamic} electron correlation in atoms and molecules.\cite{Hirata:2014gk}
Efficient implementations of MP2 may be achieved via techniques that factorize the two-electron integrals via density fitting (DF),\cite{Whitten1973p4496,Dunlap1979p3396} or Cholesky decomposition.\cite{Beebe1977p683,Koch2003p9481,Aquilante2007p194106,Aquilante2009p154107,Aquilante2011p301,Higham:2009eo}
Due to lower storage requirements, integral factorization techniques significantly reduce the cost of MP2 calculations and easily permit to target systems with 2000--3000 basis functions.\cite{Werner2003p8149}
Linear scaling\cite{ayala1999linear,Schutz1999Low-order,Werner2003p8149,doser2009linear} and stochastic\cite{willow2012stochastic,willow2013convergence,neuhauser2012expeditious} implementations of MP2 can further reduce the asymptotic computational scaling of MP2 from ${\cal O}(N^5)$ to ${\cal O}(N)$, where $N$ is the number of basis functions.

However, when MP2 is applied to study open-shell species, the buildup of \textit{static} correlation due to near-degenerate excited configurations can lead to the divergence of the correlation energy.
In this case, it is necessary to use a \textit{multireference} generalization of perturbation theory (MRPT) that can handle both \textit{dynamic} and \textit{static} correlation effects.
In practice, the distinction between \textit{dynamic} and \textit{static} correlation is enforced by dividing the full configuration interaction space into a reference space and its orthogonal complement.
The reference space consists of determinants generated by varying the occupation of the close-lying active orbitals, and consequently captures static correlation effects.
Numerous multireference perturbation theories have been proposed,\cite{Roos1992p1218,Hirao1992p374,Davidson1994p3672,Malrieu2001p10252,Mukherjee2005p134105,Hoffmann:2009ih,Evangelista2009p4728} many of which have been conveniently reviewed and compared in Refs. \citenum{Mukherjee2005p134105} and \citenum{Hoffmann:2009ih}.

A troubling aspect of several multireference perturbation theories is the well-known intruder-state problem.\cite{Paldus:1993dx} Intruder states are encountered when determinants that lie within the reference space become near-degenerate with determinants that lie in the orthogonal complement.
In perturbative theories, intruders lead to divergences in the first-order amplitudes, and the corresponding potential energy curves show characteristic spikes.\cite{Malrieu1987p4930,Piecuch2000p052506,Piecuch2000p757}  
A popular solution to remove intruders is shifting the energy denominators.\cite{Roos1995p215}
However, level shifting can significantly affect computed spectroscopic constants\cite{Camacho:2009cr} and the order of electronic states.\cite{Camacho:2010co}
In second-order \textit{n}-electron valence state perturbation theory (NEVPT2),\cite{Malrieu2001p10252,Malrieu2002p9138,Cimiraglia2007p743} intruders are removed by using Dyall's modified zeroth-order Hamiltonian.\cite{Dyall1995p4909}
Nevertheless, Zgid et al.\cite{Chan2009p194107} noticed that if the three- and four-particle density cumulants are approximated then ``false intruders'' may also appear in NEVPT2.

The importance of the intruder-state problem is not limited to multireference perturbation theories.
In the case of multireference coupled cluster theories
(MRCC)\cite{Monkhorst1981p1668,Mukherjee1998p157,Mukherjee1999p6171,Pittner1999p10275,Hanrath2005p084102,Kong2009p114101,Nooijen2011p214116,Datta2012p204107,Nooijen2014p081102,Hoffmann2012p014108}
and other nonperturbative theories of dynamical correlation,\cite{Yanai2006p194106,Yanai2007p104107,Neuscamman:2010dz} intruders cause numerical instability problems.
In this case, however, it is more appropriate talk of \textit{intruder solutions}, which arise from existence of multiple solutions to the MRCC equations.\cite{Piecuch2000p052506} 
Unfortunately, it is still not clear whether or not traditional techniques used to remove intruders in MRPT can be extended to the case of nonperturbative multireference methods.
Therefore, finding a solution to the problem of intruders in MRPT might also shed light on how to create highly-accurate multireference approaches that are numerically stable.

Recently, we have proposed the driven similarity renormalization group (DSRG),\cite{Evangelista2014p054109} a many-body formalism inspired by flow renormalization group methods.\cite{Kehrein2010book, wegner1994flow, tsukiyama2011medium, tsukiyama2012medium, hergert2014ab, jurgenson2009evolution, bogner2007similarity}
The DSRG was used to formulate a theory of dynamic electron correlation that is free from divergences due to vanishing denominators.
In the unitary DSRG ansatz, the bare Hamiltonian $(\hat{H})$ is progressively brought to a block-diagonal form (renormalized) via a continuous unitary transformation
$[\hat{U}(s)]$ controlled by the so-called flow variable $s$:
\begin{align}
\label{eq:dsrg_trans}
    \hat{H} \rightarrow \bar{H}(s) = \hat{U}(s) \hat{H} \hat{U}^{\dagger}(s), \quad s \in [0, \infty) .
\end{align}
In the limit $s\rightarrow \infty$ the DSRG unitary operator $\hat{U} (s)$ is required to block-diagonalize the Hamiltonian.
More specifically, if we indicate the non-diagonal part of $\bar{H}(s)$ with $[\bar{H}(s)]_{\rm N}$,\cite{Kutzelnigg2010p299,Kutzelnigg2009p3858} then we require that in the limit of $s$ that goes to infinity, the DSRG transformation must zero the nondiagonal parts of $\bar{H}(s)$, that is $\lim_{s\rightarrow \infty} [\bar{H}(s)]_{\rm N}$ = 0.
For intermediate values of $s$, the DSRG transformation achieves a partial block-diagonalization of the Hamiltonian, leaving states that differ in
energy by less than the energy cutoff $\Lambda = s^{-1/2}$ mostly unchanged.\cite{Wilson1994p4214,Wegner2000p133,White2002p7472,Kehrein2010book}
Consequently, in the DSRG the mixing of reference-space determinants with close-lying determinants in the orthogonal complement is suppressed and intruder states are avoided.

Another distinctive aspect of the DSRG is that it employs a Fock-space many-body formalism,\cite{Lindgren1978p33,Bartlett1996p2652} such that Eq.~\eqref{eq:dsrg_trans} should be interpreted as a set of operator equations.
Nooijen and coworkers\cite{Nooijen2011p214116} recently pointed out that a many-body formulation of multireference theories is advantageous because it
removes the need to orthogonalize the excitation manifold.  The orthogonalization step is often a bottleneck that prevents computations with large active spaces.
For example, in a study of the complete active space perturbation theory (CASPT2)\cite{roos1982simple,Pulay2011p3273,Roos1990p5483,Roos1992p1218}
coupled with the density matrix renormalization group (DMRG),\cite{White1992p2863,Chan2002p4462,wouters2014density,olivares2015ab} Yanai and Kurashige
\cite{kurashige2011second} found that the perturbation theory is limited to approximately 30 active orbitals per irreducible representation due to the
required diagonalization of the overlap metric between internally-contracted configurations.

In a previous work,\cite{Li:2015iz} we formally extended the DSRG to multireference cases (MR-DSRG) by employing the generalized Wick theorem of Mukherjee and Kutzelnigg.\cite{Mukherjee1997p432}
To study the viability of the MR-DSRG approach we performed a perturbative analysis and derived a second-order MR-DSRG perturbation theory (DSRG-MRPT2).
The DSRG-MRPT2 energy and amplitude equations are surprisingly simple and lead to a computational approach that requires only the the two- and three-body cumulants of the reference wave function.
Benchmark computations on small systems (HF, N$_2$, and \textit{p}-benzyne) showed that the DSRG-MRPT2 has an accuracy comparable to that of other second-order MRPTs.
The DSRG-MRPT2 method avoids the intruder-state problem without the use of level-shifting or increasing the size of the active space, and  in addition, it is rigorously size consistent,\cite{Pople1976p1,Pople1978p545} and thus applicable to large systems.

The present work focuses on the efficient implementation of the DSRG-MRPT2 theory to extend its applicability to chemically interesting systems.
We carefully analyze each energy contribution, and realize the possibility to factorize some terms by taking advantage of the structure of the one-particle and one-hole density matrix. 
For an active space of fixed size, the improved algorithm is dominated by terms have the same computational cost of single-reference second-order M{\o}ller--Plesset perturbation theory.
The simplicity of the DSRG-MRPT2 equations allows us to utilize common integral factorization techniques,\cite{Weigend:2009bh} including density
fitting and Cholesky decomposition, to reduce the memory and disk requirements.
In addition to MP2, various electronic structure methods have benefited from these integral factorization tactics.\cite{Hattig2000p5154,Werner2003p8149,Aquilante2008p024113,Sherrill2010p184111,Sherrill2013p2687,Krylov2013p134105,Werner2013p104104}
For instance, the Cholesky-decomposed CASPT2 has been applied to systems with up to 1500 basis functions\cite{Aquilante2008p694,Lindh2010p747} and the density-fitted NEVPT2 has been used in applications with up to 2000 basis functions.\cite{ORCA,Malrieu2002p9138}

This paper proceeds as follow.
In Sec.~II, we start with an overview of the DSRG-MRPT2 theory and integral factorization techniques.
Then, in Sec.~III we analyze the computational complexity of each energy term and detail our current implementation.
Section V presents applications of DSRG-MRPT2 to evaluate the singlet-triplet splittings of naphthynes.
Finally, we discuss future developments of the DSRG-MRPT2.

\section{Theory}
\subsection{The MR-DSRG formalism}
In this section we briefly summarize the MR-DSRG approach.\cite{Li:2015iz}
We assume that the reference is defined by a set of spin orbitals $\{ \phi_p \}$ partitioned into core ($\bf C$), active ($\bf A$), and virtual ($\bf V$) subsets of size $N_{\bf C}$, $N_{\bf A}$, and $N_{\bf V}$, respectively.
Core orbitals are designated by indices $m, n$, active orbitals by indices $u, v, w, x, y, z$, and virtual orbitals by indices $e, f$.
We also introduce two composite orbital subsets: hole ($\bf H = C \cup A$) and particle ($\bf P = V \cup A$) of dimension $N_{\bf H} = N_{\bf C} + N_{\bf A}$ and $N_{\bf P} = N_{\bf V} + N_{\bf A}$, respectively.
Orbitals belonging to hole set are associated with the labels $i, j, k, l$, while particle orbitals are labeled with $a, b, c, d$.  
General orbitals (hole or particle) are labeled as $p, q, r, s$.  

We consider the case of a complete active space (CAS) self-consistent field (CASSCF) or a CAS configuration interaction (CASCI) reference wave
function $\mref$ obtained by doubly occupying the core orbitals and distributing a given number of active electrons ($n_{\text{act}}$) in the active orbitals [CAS($n_{\text{act}}$, $N_{\bf A}$)].
The reference $\mref$ defines the Fermi vacuum with respect to which all operators are normal ordered according to Mukherjee and Kutzelnigg's generalized Wick theorem.\cite{Mukherjee1997p561,Mukherjee1997p432,Mahapatra1998p163,Shamasundar2009p174109,Mukherjee2010p234107,Mukherjee2013p62}
From the reference wave function we also extract the one-particle density matrix ($\density{p}{q}$) as well as the two- and three-body cumulants ($\tens{\lambda}{pq}{rs}$, $\tens{\lambda}{pqr}{stu}$),\cite{Mukherjee1997p432,kutzelnigg1999cumulant,mazziotti2011two} defined as:
\begin{align}
\density{p}{q} &= \braket{\mref | \cop{p} \aop{q} | \mref}, \\
\tens{\lambda}{uv}{xy} &= \tens{\gamma}{uv}{xy} - \tens{\gamma}{u}{x}\tens{\gamma}{v}{y} + \tens{\gamma}{u}{y}\tens{\gamma}{v}{x},\\
\tens{\lambda}{uvw}{xyz} &= \tens{\gamma}{uvw}{xyz} - \sum_\pi (-1)^\pi \tens{\gamma}{u}{x}\tens{\lambda}{vw}{yz} - \det(\tens{\gamma}{u}{x}
\tens{\gamma}{v}{y} \tens{\gamma}{w}{z}),
\end{align}
where $\det(\cdot)$ indicates the sum of all permutations of lower labels with a sign factor corresponding to the parity of permutations and $\sum (-1)^\pi$ indicates a sum over all permutations of the lower and upper labels with a sign factor corresponding to the parity of a given permutation.  
Note that for a CASSCF/CASCI reference the cumulant are null unless all indices belong to the active space.
For convenience we also define the one-body cumulant as $\tens{\lambda}{u}{v} = \tens{\gamma}{u}{v}$, with $u,v \in \mathbf{A}$.
The MR-DSRG equations for the amplitude and energy [$E(s)$] are given by:
\begin{align}
\label{eq:dsrg_energy}
E(s) &= \braket{\Phi | \bar{H}(s) | \Phi}, \\
\label{eq:dsrg}
[\bar{H} (s)]_{\rm N} &= \hat{R}(s),
\end{align}
where $\hat{R}(s)$ is the source operator, a $s$-dependent Hermitian operator that drives the transformation of the Hamiltonian.
Thus, the unitary operator, $\hat{U} (s)$, is implicitly defined by $\hat{R}(s)$. 
The unitary operator $\hat{U}(s)$ that controls the DSRG transformation is expressed as the exponential of an anti-Hermitian operator $\hat{A}(s)$, that is, $\hat{U}(s) = \exp[\hat{A}(s)]$. 
The operator $\hat{A}(s)$ is conveniently expressed in terms of the coupled cluster excitation operator $\hat{T}(s)$, so that $\hat{A}(s) = \hat{T}(s) - \hat{T}^{\dag}(s)$.
Note that internal amplitudes that involve only active-orbital indices are excluded from $\hat{T}(s)$, that is $\tens{t}{uv\dots}{xy\dots}(s) = 0$ $\forall u, v, x, y \dots \in \textbf{A}$.

\subsection{The DSRG-MRPT2 method}
The starting point of the DSRG-MRPT2 approach is the partitioning of the normal-ordered Hamiltonian into a zeroth-order part [$\hat{H}^{(0)}$] plus a first-order perturbation [$\hat{H}^{(1)}$].
The zeroth-order Hamiltonian is chosen to contain the reference energy ($E_0$) and the diagonal block of the one-body operator [$\hat{F}^{(0)}$]:\cite{Li:2015iz}
\begin{align}
    \hat{H}^{(0)} &= E_0 + \hat{F}^{(0)}, \\
        \hat{F}^{(0)} &= \sum_{p} \dfock{p} \no{\sqop{p}{p}},
\end{align}
where the orbital energies $\dfock{p} = \tens{f}{p}{p}$ are the diagonal elements of the generalized Fock matrix:
\begin{equation} \label{eq:gen_fock}
    \tens{f}{p}{q} = \tens{h}{p}{q} + \sum_{rs} \tens{v}{pr}{qs} \density{r}{s}.
\end{equation}
The quantities $\tens{h}{p}{q} = \braket{\phi_{p}|\hat{h}|\phi_q}$ and $\tens{v}{pq}{rs}= \braket{\phi_{p}\phi_{q}||\phi_{r}\phi_{s}}$ are respectively one-electron and antisymmetrized two-electron integrals in the molecular orbital basis.

As is the case for other perturbation theories, we find it advantageous to formulate the \methodname in a basis of semicanonical molecular orbitals\cite{Handy1989p185} so that the core, active, and virtual blocks of the generalized Fock matrix are diagonal.
This choice implies that $\hat{F}^{(1)}$ only contains contributions from the off-diagonal blocks of the Fock matrix.

The DSRG-MRPT2 equations may be obtained from Eqs.\eqref{eq:dsrg_energy} and \eqref{eq:dsrg} by performing a order-by-order expansion.\cite{Bartlett2009book}
The zeroth-, first-, and second-order energy expressions are given by:\cite{Li:2015iz}
\begin{align}
    E^{(0)} (s) &= E_0, \label{eq:Ezero}\\
    E^{(1)} (s) &= 0, \label{eq:Efirst}\\
    E^{(2)}(s) &= \braket{[\tilde{H}^{(1)} (s), \hat{T}^{(1)} (s)]}, \label{eq:Esecond}
\end{align}
where $\tilde{H}^{(1)}$ is an effective first-order Hamiltonian with modified non-diagonal components:
\begin{equation}
\label{eq:effective_h}
\tilde{H}^{(1)}(s) = \hat{H}^{(1)}(s) + [\hat{R}^{(1)} (s)]_{\rm N},
\end{equation}
while the diagonal components of $\tilde{H}^{(1)}$ are identical to those of $\hat{H}^{(1)}$.

A first-order expansion of the MR-DSRG amplitude equations leads to the equation:
\begin{equation}
\label{eq:amplitudes}
[\hat{H}^{(1)}]_{\rm N} + [\hat{H}^{(0)}, \hat{T}^{(1)}]_{\rm N} = [\hat{R}^{(1)}(s)]_{\rm N},
\end{equation}
from which explicit equations for the the first-order amplitudes can be derived:\cite{Li:2015iz}
\begin{align}
    \tens{t}{a}{i,(1)} (s) &= [ \tens{f}{a}{i,(1)} + \sum\limits_{ux}^{\mathbf{A}} \tens{\Delta}{u}{x} \tens{t}{ax}{iu,(1)} (s) \density{x}{u} ] \frac{1 - e^{-s(\tens{\Delta}{a}{i})^2}}{\tens{\Delta}{a}{i}}, \label{eq:T1amp} \\
    \tens{t}{ab}{ij,(1)} (s) &= \tens{v}{ab}{ij,(1)} \frac{1 - e^{-s(\tens{\Delta}{ab}{ij})^2}}{\tens{\Delta}{ab}{ij}}. \label{eq:T2amp}
\end{align}
Here we have introduced the M{\o}ller--Plesset denominators $\tens{\Delta}{ab\cdots}{ij\cdots}$, defined as $\tens{\Delta}{ab\cdots}{ij\cdots} = \dfock{i} + \dfock{j} + \ldots - \dfock{a} - \dfock{b} - \ldots$.
In the derivation of Eqs.~\eqref{eq:T1amp} and \eqref{eq:T2amp} we used the source operator introduced in Ref.~\citenum{Evangelista2014p054109}, which is designed to reproduce the energy of the second-order similarity renormalization group.\cite{hergert2016medium}

Once the first-order amplitudes are solved, the second-order energy $E^{(2)}(s)$ can be obtained via an efficient non-iterative procedure that requires at most three-body density cumulants.
For convenience, we list all \methodname energy contributions in Table~\ref{tab:energy}.
These quantities are expressed in terms of the modified first-order Fock matrix matrix elements:
\begin{align}
    \tilde{f}_{a}^{i, (1)} (s) =& f_{a}^{i, (1)} [1 + e^{-s(\Delta_{a}^{i})^2}] \notag \\
    &+ [\sum_{ux}\, \tens{\Delta}{u}{x} \tens{t}{ax}{iu,(1)} (s) \density{x}{u}] e^{-s(\Delta_{a}^{i})^2} \label{eq:f_tilde},
\end{align}
the modified two-electron integrals:
\begin{align}
    \tilde{v}_{ab}^{ij, (1)} (s) =& v_{ab}^{ij, (1)} [1 + e^{-s(\Delta_{ab}^{ij})^2}], \label{eq:v_tilde}
\end{align}
the one-particle and one-hole density matrix elements $(\density{p}{q}, \cdensity{p}{q} = \kro{p}{q} - \density{p}{q})$, and the two- and three-body density cumulants $(\cumulant{uv}{xy}, \cumulant{uvw}{xyz})$ of the reference $\mref$.
Eqs.~\eqref{eq:T1amp}--\eqref{eq:v_tilde} and the equations reported in Table~\ref{tab:energy} define the \methodname method.

To highlight the mechanism by which the \methodname avoids intruders, we perform a Maclaurin expansion of the first-order amplitudes as
a function of the energy denominators.
For example, the $t_2$ amplitude [Eq.~\eqref{eq:T2amp}] can be rewritten as:
\begin{align}
    \tens{t}{ab}{ij,(1)} (s) &= \tens{v}{ab}{ij,(1)} \left( s \tens{\Delta}{ab}{ij} + {\cal O}[s^{3/2} (\tens{\Delta}{ab}{ij})^3 ] \right),
\end{align}
which approaches zero in the limit of $|\tens{\Delta}{ab}{ij}| \rightarrow 0$.
Thus for finite values of $s$, the second-order energy, $E^{(2)}(s)$, is well-behaved and free from divergences due to small energy denominators.
One of the drawbacks of the \methodname renormalization procedure is that the final energy shows a dependence on the value of $s$ used in a computation.
In our previous work,\cite{Li:2015iz} we analyzed the $s$-dependence of the DSRG-MRPT2 energy and found that the range $s \in [0.1, 1.0]$ \sunit gives the best agreement with full configuration interaction results.
Values of $s$ that fall out of this ``Goldilocks zone'' either lead to recovering too little correlation energy (when $s \ll 0.1$) or expose the theory to the intruder state problem (when $s \gg 1$).

\begin{table}[!t]
\ifpreprint
\renewcommand{\arraystretch}{0.95}
\else
\renewcommand{\arraystretch}{1.5}
\fi
  \caption{\methodname second-order energy expressions. The Einstein convention for the summation over repeated indices is employed. Asymptotic scalings are given in big $\cal O$ notation from a straightforward tensor-index analysis.}
   \label{tab:energy}
    {
      \begin{tabular*}{\columnwidth}{@{\extracolsep{\stretch{1}}}*{1}{c}*{2}{l}@{}}
       \hline
       \hline
       Term & Energy Expression & Cost \\
       \hline
       \multicolumn{3}{c}{$\braket{[\tilde{F}^{(1)} (s),\hat{T}_{1}^{(1)} (s)]}$ \rule{0pt}{4ex}} \\
       \hline
       A & $+ \tens{\tilde{f}}{j}{b,(1)} (s) \tens{t}{a}{i,(1)} (s)
       \density{j}{i}\cdensity{a}{b}$ & $N_{\textrm{P}}^2 N_{\textrm{H}}^2$ \\[3pt]
       \multicolumn{3}{c}{$\braket{[\tilde{V}^{(1)} (s), \hat{T}_1^{(1)} (s)]}$}\\
	              \hline
       B & $+ \frac{1}{2} \tens{\tilde{v}}{xy}{ev,(1)} (s) \tens{t}{e}{u,(1)} (s) \cumulant{xy}{uv}$ & $N_{\textrm{A}}^4 N_{\textrm{V}}$ \\
       C & $- \frac{1}{2} \tens{\tilde{v}}{my}{uv,(1)} (s) \tens{t}{x}{m,(1)} (s) \cumulant{xy}{uv}$ & $N_{\textrm{A}}^4 N_{\textrm{C}}$ \\[3pt]
       \multicolumn{3}{c}{$\braket{[\tilde{F}^{(1)} (s), \hat{T}_2^{(1)} (s)]}$}\\
       \hline       
       D& $+ \frac{1}{2} \tens{\tilde{f}}{x}{e,(1)} (s) \tens{t}{ey}{uv,(1)} (s)
       \cumulant{xy}{uv}$ & $N_{\textrm{A}}^4 N_{\textrm{V}}$ \\
       E & $- \frac{1}{2} \tens{\tilde{f}}{m}{v,(1)} (s) \tens{t}{xy}{um,(1)} (s) \cumulant{xy}{uv}$ & $N_{\textrm{A}}^4 N_{\textrm{C}}$ \\[3pt]
       \multicolumn{3}{c}{$\braket{[\tilde{V}^{(1)} (s),\hat{T}_{2}^{(1)} (s)]}$}\\
       \hline
       F & $+ \frac{1}{4} \tens{\tilde{v}}{kl}{cd,(1)}(s)
       \tens{t}{ab}{ij,(1)}(s) \density{k}{i} \density{l}{j} \cdensity{a}{c} \cdensity{b}{d}$ & $N_{\textrm{P}}^3N_{\textrm{H}}^2$ \\
       G & $+ \frac{1}{8} \tens{\tilde{v}}{xy}{cd,(1)}(s) \tens{t}{ab}{uv,(1)}(s) \cdensity{a}{c} \cdensity{b}{d} \cumulant{xy}{uv}$ & $N_{\textrm{A}}^4N_{\textrm{P}}^2$ \\
       H & $+ \frac{1}{8} \tens{\tilde{v}}{kl}{uv,(1)}(s) \tens{t}{xy}{ij,(1)}(s) \density{k}{i} \density{l}{j} \cumulant{xy}{uv}$ & $N_{\textrm{A}}^4N_{\textrm{H}}^2$ \\
       I & $+ \tens{\tilde{v}}{jx}{bu,(1)}(s) \tens{t}{ay}{iv,(1)}(s) \density{j}{i} \cdensity{a}{b} \cumulant{xy}{uv}$ & $N_{\textrm{A}}^4N_{\textrm{P}} N_{\textrm{H}}$ \\
       J & $+ \frac{1}{4} \tens{\tilde{v}}{mz}{uv,(1)}(s) \tens{t}{xy}{mw,(1)}(s) \cumulant{xyz}{uvw}$ & $N_{\textrm{A}}^6N_{\textrm{C}}$ \\
       K & $+ \frac{1}{4} \tens{\tilde{v}}{xy}{we,(1)}(s) \tens{t}{ez}{uv,(1)}(s) \cumulant{xyz}{uvw}$ & $N_{\textrm{A}}^6N_{\textrm{V}}$ \\
       \hline
       \hline
      \end{tabular*}
    }
\end{table}

\subsection{Integral factorizations}
The simple structure of the MR-DSRG amplitude and energy equations (Table~\ref{tab:energy}) allows the use of integral factorization techniques such as DF and/or Cholesky decomposition to improve the efficiency of the \methodname.  Integral factorization techniques seek to approximate the electron repulsion integrals as a contraction of two three-index tensors.  The two-electron integrals written in chemist notation can be factorized as:
\begin{equation}
\label{eq:int_fact}
    (pq | rs) \approx \sum_Q^M B^Q_{pq} B^{Q}_{rs},
\end{equation}
where $M$ is the size of the auxiliary basis set $\{ \chi_{P}({\bf r}) \}$.  
In the DF approach, the factors $B^Q_{pq}$ are given by:\cite{kendall1997impact}
\begin{align}
    B_{pq}^{Q} = \sum_{P} (pq | P) [{\bf J}^{-1/2}]_{PQ},
\end{align}
where $(pq | P)$ and $J_{PQ}$ are three- and two-center integrals defined as:
\begin{align}
    (pq | P) &= \int {\rm d} {\bf r}_1 \int {\rm d} {\bf r}_2 \, \phi_{p}({\bf r}_1) \phi_{q}({\bf r}_1) \, r_{12}^{-1} \, \chi_{P}({\bf r}_2),\\
    J_{PQ} &= \int {\rm d} {\bf r}_1 \int {\rm d} {\bf r}_2 \, \chi_{P}({\bf r}_1) \, r_{12}^{-1} \, \chi_{Q}({\bf r}_2).
\end{align}
In this work we evaluate the \methodname energy using the resolution of the identity (RI) basis sets of Weigend and co-workers.\cite{weigend2002efficient}
We note, however, that there is no consensus on the most appropriate auxiliary basis set for multireference perturbation theories.

In the CD approach, the Cholesky factors $B^Q_{pq}$ are obtained directly via decomposition of the four-index two-electron integrals.\cite{Aquilante2011p301}
The CD approach generates the auxiliary basis set by a numerical Cholesky decomposition.\cite{golub2012matrix}
As such, CD is sometimes referred to \abinitio density fitting.\cite{Aquilante2007p194106,Aquilante2009p154107}
The upper bound of the summation $M$ in Eq.~\eqref{eq:int_fact} is determined by a CD threshold, which measures the error introduced by the Cholesky decomposition.\cite{Aquilante2009p154107,Aquilante2011p301}

\section{Implementation}
An efficient implementation of the \methodname is achieved by taking advantage of the structure of the density matrices and integral factorization.
In most practically relevant cases, the number of active orbitals is negligible compared to the number of core and virtual orbitals, that is we may assume that:
\begin{equation}
N_{\bf A} \ll N_{\bf C} < N_{\bf V}.
\end{equation}
Under this assumption, the most expensive term in the evaluation of the \methodname energy is term F of Table \ref{tab:energy}.
This term originates from the contraction $\braket{[\tilde{V}^{(1)}(s),\hat{T}^{(1)}_2(s)]}$ and is given by:
\begin{equation} \label{eq:expensive_term}
\mathrm{F} =  \frac{1}{4} \sum_{ijkl}^{\mathbf{H}}  \sum_{abcd}^{\mathbf{P}} \tens{\rv}{kl}{cd,(1)}(s) \, \tens{t}{ab}{ij,(1)}(s) \, \density{k}{i} \density{l}{j} \cdensity{a}{c} \cdensity{b}{d}.
\end{equation}
The computational cost required to evaluate term F scales formally as ${\cal O}(N_{\bf P}^4 N_{\bf H}^4)$, but can be reduced to ${\cal O}(N_{\bf P}^3 N_{\bf H}^2)$ via factorization into intermediate tensors.

\begin{figure}[t]
\includegraphics[width=3.0in]{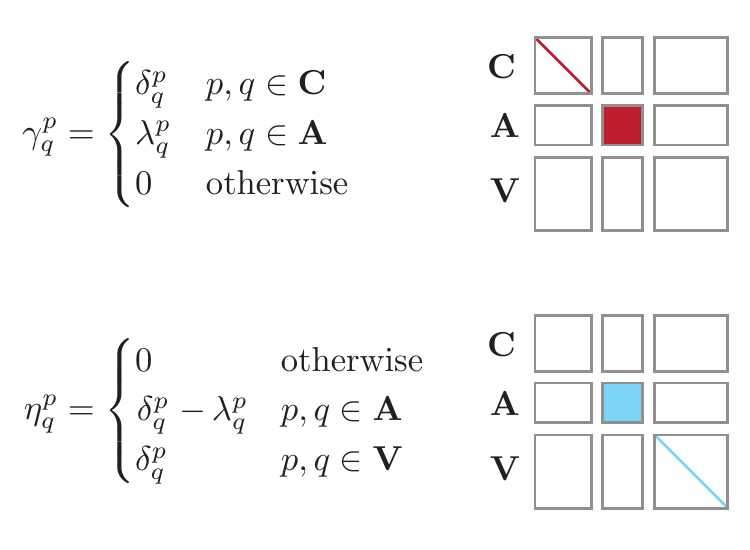}
\caption{The structures of one-particle density matrix $\density{p}{q}$ and one-hole density matrix $\cdensity{p}{q}$  Non-zero elements of the one-particle and one-hole density matrices are indicated respectively in red and blue.} 
\label{fig:densities}
\end{figure}

For a CASSCF/CASCI reference, we can reduce the cost of evaluating term F by taking advantage of the structure of the one-particle and one-hole density matrices.
As illustrated in Fig.~\ref{fig:densities}, $\tens{\gamma}{q}{p}$ is diagonal in the core-core block and in the active-active block it is equal to the one-body cumulant $\tens{\lambda}{q}{p}$.
Upon explicit replacement of the one-body density and hole density matrices into Eq.~\eqref{eq:expensive_term} we obtain eight contributions (F1--F8) that are reported in Table~\ref{eq:diagrams}.
Each term is also represented as a diagram in which one or more lines pass through a one-particle (red circle) or one-hole (blue circle) vertex.
The most expensive contributions to term F [Eq.~\eqref{eq:expensive_term}] is diagram F1, which has a computation scaling of ${\cal O}(N_{\bf V}^2 N_{\bf C}^2)$, followed by F2 and F3, which scale as ${\cal O}(N_{\bf V}^2 N_{\bf A} N_{\bf C})$ and ${\cal O}(N_{\bf V} N_{\bf A} N_{\bf C}^2)$, respectively.
The remaining diagrams shown in Table~\ref{eq:diagrams} (F4--F8) carry at least two active indices and are significantly less expensive to evaluate.

\begin{table}[t]
\centering
\ifpreprint
\renewcommand{\arraystretch}{0.85}
\else
\renewcommand{\arraystretch}{1.2}
\fi
\caption{\methodname energy terms that arise from diagram F after taking into account the block structure of the one-hole and one-particle density matrices.  Contractions involving the one-particle density matrix ($\tens{\gamma}{j}{i}$) and hole indices are indicated with a red circle, while contractions of the one-hole density matrix ($\tens{\eta}{b}{a}$) and particle indices are indicated with a blue circle.}
\label{tab:daigrams}
\begin{tabular*}{\columnwidth}{@{\extracolsep{\stretch{1}}}*{2}{c}l@{}}
\hline
\hline
Term &  Diagram &  Expression \\
\hline
F1 & \raisebox{-.5\height}
{\includegraphics[height=0.4in]{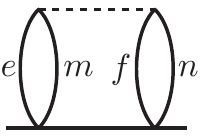}} & $\frac{1}{4} \sum \limits_{mnef} \tens{\rv}{mn}{ef,(1)} (s) \tens{t}{ef}{mn,(1)} (s)$ \\[14pt]
F2 & \raisebox{-.5\height}{\includegraphics[height=0.4in]{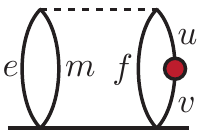}} & $\frac{1}{2} \sum \limits_{mefuv} \tens{\rv}{mu}{ef,(1)} (s) \tens{t}{ef}{mv,(1)} (s) \density{u}{v}$ \\[14pt]
F3 & \raisebox{-.5\height}{\includegraphics[height=0.4in]{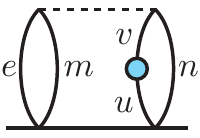}} & $\frac{1}{2} \sum \limits_{mneuv} \tens{\rv}{mn}{ev,(1)} (s) \tens{t}{eu}{mn,(1)} (s) \cdensity{u}{v}$ \\[14pt]
F4 & \raisebox{-.5\height}{\includegraphics[height=0.4in]{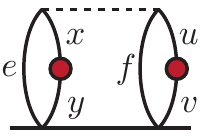}} & $\frac{1}{4} \sum \limits_{ef} \sum \limits_{uvxy} \tens{\rv}{xu}{ef,(1)} (s) \tens{t}{ef}{yv,(1)} (s) \density{x}{y} \density{u}{v}$ \\[14pt]
F5 & \raisebox{-.5\height}{\includegraphics[height=0.4in]{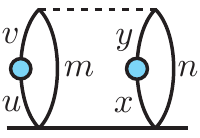}} & $\frac{1}{4} \sum \limits_{mn} \sum \limits_{uvxy} \tens{\rv}{mn}{vy,(1)} (s) \tens{t}{ux}{mn,(1)} (s) \cdensity{u}{v} \cdensity{x}{y}$ \\[14pt]
F6 & \raisebox{-.5\height}{\includegraphics[height=0.4in]{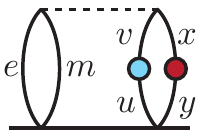}} & $\sum \limits_{me} \sum \limits_{uvxy} \tens{\rv}{mx}{ve,(1)} (s) \tens{t}{ue}{my,(1)} (s) \density{x}{y} \cdensity{u}{v}$ \\[14pt]
F7 & \raisebox{-.5\height}{\includegraphics[height=0.4in]{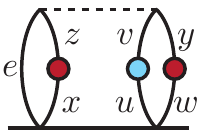}} & $\frac{1}{2} \sum \limits_{wxyz} \sum \limits_{euv} \tens{\rv}{yz}{ve,(1)} (s) \tens{t}{ue}{wx,(1)} (s) \density{y}{w} \density{z}{x} \cdensity{u}{v}$ \\[14pt]
F8 & \raisebox{-.5\height}{\includegraphics[height=0.4in]{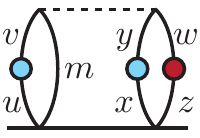}} & $\frac{1}{2} \sum \limits_{mwz} \sum \limits_{uvxy} \tens{\rv}{mw}{vy,(1)} (s) \tens{t}{ux}{mz,(1)} (s) \density{w}{z} \cdensity{u}{v} \cdensity{x}{y}$ \\[14pt]
\hline
& \raisebox{-.5\height}{\includegraphics[height=0.45in]{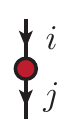}} & $\density{i}{j}$ \\
& \raisebox{-.5\height}{\includegraphics[height=0.45in]{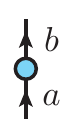}} & $\cdensity{a}{b}$ \\
\hline
\hline
\end{tabular*}
\label{eq:diagrams}
\end{table}

\begin{figure}
\includegraphics[width=2.5in]{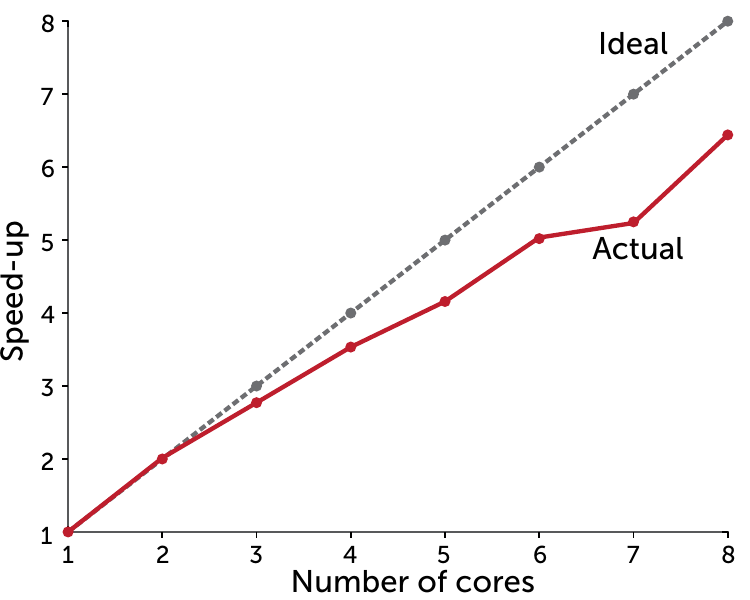}
\caption{The scaling of diagram F1 of Table~\ref{eq:diagrams} for $(2,3)$-naphthyne using a cc-pVTZ basis set.  The speed up is determined as $\frac{S(1)}{S(N)}$ where $S(i)$ is the total time required to evaluate this term using $i$ threads.  Results are for up to 8 threads on an Intel Xeon E5-2650 v2 processor.}
\label{fig:scaling_expen}
\end{figure}

Diagram F1 may be written in a form that is reminiscent of the MP2 correlation energy:
\begin{equation}
\frac{1}{4} \sum_{mn}^{\mathbf{C}}\sum_{ef}^{\mathbf{V}}|\tens{v}{mn}{ef}|^2 \,
\frac{1 - e^{-2s (\tens{\Delta}{ef}{mn})^2}}{\tens{\Delta}{ef}{mn}}.
\label{eq:expensive}
\end{equation}
Eq.~\eqref{eq:expensive} can be implemented in an efficient way by an outer loop over pairs of occupied orbitals $m$ and $n$.  
For each pair $(m,n)$ we compute all the antisymmetrized two electron integrals $\{\tens{v}{mn}{ef} , \forall e,f\}$ using the DF or CD factors.
The integrals squared are then contracted with the renormalized denominators $[1- \mathrm{e}^{-2s (\Delta_{ef}^{mn})^2}]/ \Delta_{ef}^{mn}$ through a dot-product operation to give a pair energy for every $m$ and $n$.\cite{Bernholdt1996p477}
The loop over the $(m,n)$ pairs is parallelized using OpenMP for shared memory architectures.
The scaling of the implementation of Eq.~\eqref{eq:expensive} on a eight-core processor is demonstrated in Fig.~\ref{fig:scaling_expen}.
Our implementation is also optimized for the evaluation of diagrams F2 and F3 so that no storage of large four-index intermediate quantities is necessary.

\begin{figure}
\includegraphics[width=3.35in]{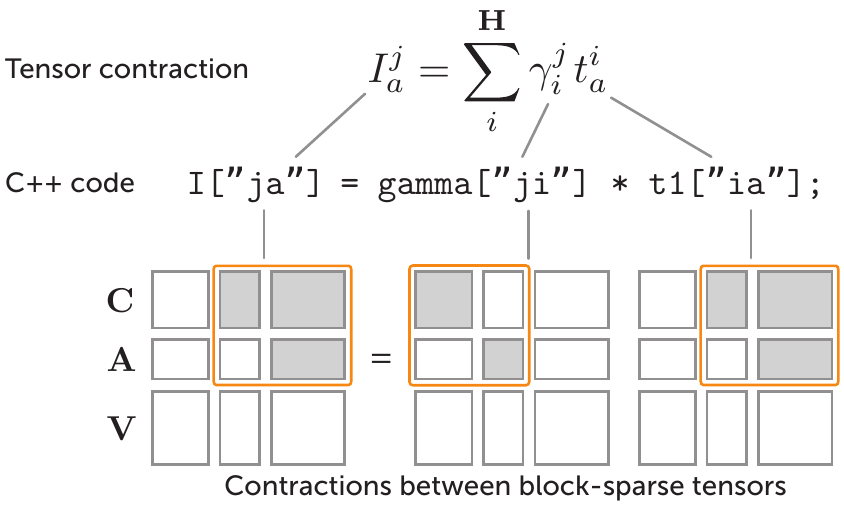}
\caption{This figure illustrates the ability of the \ambit{} tensor library to deal with block-sparse tensors that span composite orbital space.  The tensor contraction shown at the top involves a summation over the index $j$ that spans the generalized orbitals space ($\mathbf{H}$), which is the union of the core ($\mathbf{C}$) and active ($\mathbf{A}$) orbitals.
The tensors $\gamma_i^j$ and $t_a^i$ are defined over subsets (shown in orange) of the full orbital indices and are block sparse.  For example, the $\mathbf{A}$-$\mathbf{A}$ block of $t_a^i$ is zero because internal amplitudes are not defined in the \methodname{}.
\ambit{} allows to write contractions over block-sparse tensors as contractions over composite index tensors, thus, reducing the number of equations required to include in the source code.}
\label{fig:ambit_sample}
\end{figure}

The \methodname equations are implemented in our code \textsc{Forte},\cite{FORTE2015} a suite of multireference methods written as a plugin to the \PSI quantum chemistry package.\cite{PSI} All tensor contractions were coded using the open-source library \ambit.\cite{AMBIT2015}
\ambit provides shared memory parallelization and performs tensor contractions using BLAS operations.
A very convenient feature of \ambit{} is its ability to deal with composite orbital spaces.
Figure~\ref{fig:ambit_sample} gives an example of a tensor contraction encountered in the DSRG-MRPT2 equations and how it is implemented via \ambit{}.
Composite spaces are defined from ``primitive'' spaces (for example, the sets of core, active, virtual MOs) and arise naturally in all multireference theories based on a CASCI/CASSCF reference.
\ambit{} is aware of composite orbital spaces and can perform contractions over block-sparse tensors.
This feature greatly simplifies the implementation of multireference theories since it allows the user to directly encode tensor contractions that involve composite orbital indices.

In summary, the following procedure was used for computing the DSRG-MRPT2 energy using DF or CD integrals:  
\begin{enumerate}
\setlength\itemsep{0pt}
\item Compute $\tens{\gamma}{p}{q}$, $\tens{\eta}{p}{q}$, $\tens{\lambda}{uv}{xy}$, and $\tens{\lambda}{uvw}{xyz}$ for the CASSCF/CASCI reference.
\item Compute the Fock matrix from $\tens{\gamma}{p}{q}$ and the DF/CD tensors.  
\item Canonicalize the core, active, and virtual MOs.
\item Form the antisymmetrized two electron integrals with at least one active index from the DF/CD tensors.
\item Transform all the density matrices, cumulants, and integrals to the semi-canonical basis.
\item Compute the second order energy terms A--E, G--K, and F4--F8 using the \ambit{} library.
\item Compute the energy terms F1--F3 with an optimized algorithm that does not require storage of four-index intermediates.
\end{enumerate}

\section{Computational Details}

\label{sec:computational_details}
\begin{figure}[th]
\ifpreprint
\includegraphics[width=0.25\columnwidth]{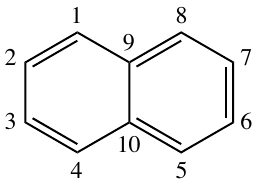}
\else
\includegraphics[width=0.5\columnwidth]{figure_4.pdf}
\fi
\caption{Naphthalene numbering scheme used in this study.
The notation ($i, j$)-naphthyne indicates that two hydrogen atoms were removed from positions $i$ and $j$.} 
\label{fig:molecules}
\end{figure}

In this work, we studied the singlet-triplet splittings ($\Delta E_{\rm ST} = E_{\rm S} - E_{\rm T}$) of ten naphthyne isomers.
Each isomer is designated as ($i, j$)-naphthyne, and it is formally obtained by removing two hydrogens from the carbons at $i$ and $j$ positions of a naphthalene.
Figure \ref{fig:molecules} shows the numbering scheme of naphthalene used in this work.

Following Ref.~\citenum{li2009energetics}, we optimized the geometry of singlet (1,3)-, (2,6)-, and (1,6)-naphthyne isomers at the CASSCF/cc-pVDZ level of theory with a CAS(4,4), CAS(2,2), and CAS(2,2) active space, respectively.
All other naphthyne isomers were optimized using Becke's three-parameter exchange\cite{becke1993density} and Lee-Yang-Parr
correlation\cite{lee1988development} (B3LYP) functional and the cc-pVDZ\cite{Dunning1989p1007} basis set.  
Unrestricted Kohn-Sham orbitals were used for both singlet and triplet states.
Geometry optimizations were performed using the \nwchem\cite{NWCHEM6} software package.

\begin{table}[H]
\centering
\ifpreprint
\renewcommand{\arraystretch}{0.85}
\else
\renewcommand{\arraystretch}{1.2}
\fi
\caption{Point group symmetries for all naphthyne isomers along with the corresponding minimal active spaces in Cotton's
ordering.\cite{cotton2008chemical}  For (1,4)-, (1,8)-, (2,3)-, and (2,7)-naphthynes, the molecules are placed in the $xz$ plane, where $z$ is the $C_2$
rotation axis. All other naphthynes are placed in the $xy$ plane.}
\label{tab:active}
\begin{tabular*}{\columnwidth}{@{\extracolsep{\stretch{1}}}*{3}{c}*{2}{r}@{}}
\hline
\hline
 & &  & \multicolumn{2}{c}{Active Space} \\
\cline{4-5}
Isomer & Sym.  & States & CAS(2,2) & CAS(12,12) \\
\hline
1,2 & $C_s$    & $^1A'$, $^3A'$ & $(2,0)$ & $(2, 10)$\\
1,3 & $C_s$    & $^1A'$, $^3A'$ & $(2,0)$ & $(2, 10)$\\
1,4 & $C_{2v}$ & $^1A_1$, $^3B_1$ & $(1,0,1,0)$ & $(1, 5, 1, 5)$\\
1,5 & $C_{2h}$ & $^1A_g$, $^3B_u$ & $(1,0,0,1)$ & $(1, 5, 5, 1)$\\
1,6 & $C_s$    & $^1A'$, $^3A'$ & $(2,0)$ & $(2, 10)$\\
1,7 & $C_s$    & $^1A'$, $^3A'$ & $(2,0)$ & $(2, 10)$\\
1,8 & $C_{2v}$ & $^1A_1$, $^3B_1$ & $(1,0,1,0)$ & $(1, 5, 1, 5)$\\
2,3 & $C_{2v}$ & $^1A_1$, $^3B_1$ & $(1,0,1,0)$ & $(1, 5, 1, 5)$\\
2,6 & $C_{2h}$ & $^1A_g$, $^3B_u$ & $(1,0,0,1)$ & $(1, 5, 5, 1)$\\
2,7 & $C_{2v}$ & $^1A_1$, $^3B_1$ & $(1,0,1,0)$ & $(1, 5, 1, 5)$\\
\hline
\hline
\end{tabular*}
\end{table}
State-specific \methodname computations used a CASCI reference.
The active spaces for different naphthyne isomers are reported in Table \ref{tab:active}.
Since at the moment we do not have access to a DF/CD CASSCF implementation, we opted for evaluating the energy of both the singlet and triplet states using restricted open-shell Hartree--Fock (ROHF) orbitals.
This choice of orbitals is certainly not optimal, and may lead to an imbalanced treatment of singlet and triplet states.
Dunning's correlation-consistent cc-pV$X$Z ($X =$ D, T, Q, 5) basis sets\cite{Dunning1989p1007,Dunning1994p2975} were used to deduce basis set effects, and the corresponding auxiliary basis sets were chosen as cc-pV$X$Z-JKFIT basis sets\cite{weigend2002fully} for ROHF computations and cc-pV$X$Z-RI basis sets\cite{weigend2002efficient,hattig2005optimization} for \methodname computations.
We used a value of $s = 0.5$ \Eht, and kept the $1s$-like orbitals on carbon atoms frozen for all \methodname computations.

\section{Results}
\label{sec:naphthynes}
\subsection{Singlet-triplet splittings of naphthyne diradicals}
In this section we will demonstrate how our efficient implementation of the DSRG-MRPT2 can be used to obtain the singlet-triplet splitting of naphthynes with fairly large basis sets.
Among arynes,\cite{wenk2003one,sanz2008recent,abe2012chemistry} the electronic structure of \textit{ortho}, \textit{meta}, and \textit{para} benzyne has been well characterized from the point of view of both experiment and theory.\cite{hoffmann1968benzynes,squires1998electronic,wenthold1998ultraviolet,Cramer:1998tz,Crawford:2001we,evangelista2007coupling,Wang:2008ix,li2009energetics}
However, in the case of naphthynes, singlet-triplet splittings have been investigated mostly by theoretical studies\cite{squires1998electronic,li2009energetics,brabec2011massively,brabec2012towards} and, to the best of our knowledge, no experimental values have been reported.
\begin{table}[H]
\centering
\begin{threeparttable}
\ifpreprint
\renewcommand{\arraystretch}{0.85}
\else
\renewcommand{\arraystretch}{1.2}
\fi
\caption{Analysis of the \methodname energy error (in \kcal) introduced by density fitting (DF) and Cholesky decomposition (CD).
Statistics were computed from the singlet-triplet splittings of the ten naphthyne isomers. 
Density fitting results were obtained using the cc-pVDZ-RI auxiliary basis set, while Cholesky vectors were generated using a threshold of 10$^{-5}$ \Eh.}
\label{tab:MAD_MAX}
\begin{tabular*}{\columnwidth}{@{\extracolsep{\stretch{1}}}cccc@{}}
\hline
\hline
Factorization & Statistics & \multicolumn{1}{c}{VDZ*\tnote{a}} & \multicolumn{1}{c}{cc-pVDZ} \\
\hline
\multirow{3}{*}{DF} &MAX\tnote{b} & 0.017 & 0.017 \\
 & MAE\tnote{c} &  0.007 & 0.007 \\
 & $\sigma$\tnote{d} &  0.005 & 0.005 \\
\hline
\multirow{3}{*}{CD} & MAX\tnote{b} & 0.003 & 0.004 \\
 & MAE\tnote{c} &  0.002 & 0.002 \\
 & $\sigma$\tnote{d} &  0.001 & 0.001 \\
\hline
\hline
\end{tabular*}
\begin{tablenotes}
\item [a] The VDZ* basis set is constructed from the cc-pVDZ basis set by removing the $p$ functions for hydrogen atoms.
\item [b] Maximum absolute error: ${\rm MAX} = \max (|\Delta_{i}|)$.
\item [c] Mean absolute error: ${\rm MAE} = \frac{1}{10} \sum_{i=1}^{10} |\Delta_{i}|$.
\item [d] Standard deviation: $\sigma = [\frac{1}{10} \sum_{i=1}^{10}(\Delta_i - \bar{\Delta})^2]^{1/2}$, where $\bar{\Delta} = \frac{1}{10} \sum_{i=1}^{10} \Delta_{i}$.
\end{tablenotes}
\end{threeparttable}
\end{table}

We first verify the accuracy of the integral factorization techniques by performing DSRG-MRPT2 computations with DF, CD, and conventional integrals.
Table~\ref{tab:MAD_MAX} reports an analysis of the errors introduced by the DF and CD approximations when applied to compute $\Delta_{\rm ST}$.  
These results shows that both approximations introduce errors that are well within chemical accuracy: the maximum absolute error for DF and CD is only $0.017$ and $0.003$ \kcal, respectively.

\newcolumntype{d}[1]{D{.}{.}{#1}}
\begin{table*}[ht!]
\begin{threeparttable}
\centering
\ifpreprint
\renewcommand{\arraystretch}{0.85}
\else
\renewcommand{\arraystretch}{1.2}
\fi
\caption{Adibatic singlet-triplet splittings ($\Delta E_{\rm ST} = E_{\rm S} - E_{\rm T}$) of naphthyne diradicals computed with the DF-\methodname approach and a variety of basis sets.  All computations utilized ROHF triplet orbitals.  Carbon 1s-like orbitals were excluded from the computations of the correlation energy.}
\label{tab:dsrg-mrpt2-ST-ccpvdz}
\ifpreprint
\begin{tabular*}{\columnwidth}{@{\extracolsep{\stretch{1}}}cc*{3}{d{3.1}}*{7}{d{2.1}}@{}}
\else
\begin{tabular*}{2\columnwidth}{@{\extracolsep{\stretch{1}}}cc*{3}{d{3.1}}*{7}{d{2.1}}@{}}
\fi
\hline
\hline
 &  & \multicolumn{10}{c}{Naphthyne Isomers} \\
 \cline{3-12}
 &  & \multicolumn{2}{c}{Group I} & \multicolumn{1}{c}{Group II} & \multicolumn{7}{c}{Group III} \\
  \cline{3-4} \cline{5-5} \cline{6-12} \vspace{-12pt} \\
Active Space & Basis & \multicolumn{1}{c}{1,2} & \multicolumn{1}{c}{2,3} & \multicolumn{1}{c}{1,3} & \multicolumn{1}{c}{1,5} & \multicolumn{1}{c}{1,6} & \multicolumn{1}{c}{1,4} & \multicolumn{1}{c}{2,7} & \multicolumn{1}{c}{2,6} & \multicolumn{1}{c}{1,7} & \multicolumn{1}{c}{1,8} \\
\hline
\multirow{4}{*}{CAS(2,2)}   
    & cc-pVDZ   & -30.2 & -24.9 & -11.7 & 1.3 & 1.4  & 6.4 & 0.9 &   1.3 & 5.2 & 3.7 \\
    & cc-pVTZ   & -33.5 & -28.3 & -14.1 & 0.3 & -0.8 & 5.3 & 0.3 & -0.8  & 4.2 & 3.0 \\
    & cc-pVQZ   & -34.4 & -29.2 & -14.4 & 0.0 & -1.1 & 5.1 & 0.1 & -1.1  & 4.1 & 2.9 \\
    & cc-pV5Z   & -34.7 & -29.5 & -14.5 & 0.0 & -1.2 & 5.1 & 0.0 & -1.1  & 4.0 & 2.9 \\
\hline
\multirow{3}{*}{CAS(12,12)}
    & cc-pVDZ       & -29.0 & -24.3 & -11.0 &   0.8& 2.0  & 5.3 &1.3& 1.3  & 5.2  & 4.3\\ 
    & cc-pVTZ       & -32.6 & -27.8 & -13.7 &  -0.4& -0.2 & 4.2 &0.6& -0.8 & 4.2  & 3.4\\
    & cc-pVQZ       & -33.6 & -28.9 & -14.2 &  -0.6& -0.5 & 4.1 &0.3& -1.0 & 4.0  & 3.3\\
\hline 
    \multicolumn{2}{c}{$c_1/c_2$\tnote{a}} & 2.6 & 2.5 & 1.6 & 1.3 & 1.0 & 1.3 &1.1 & 1.0 & 1.1 & 1.2  \\
\hline
\hline
\end{tabular*}
\begin{tablenotes}
\item [a] The ratio of CI coefficients between the two dominant determinants in a CAS(2,2).  This characteristic was used to separate the
naphthynes into three separate groups.  
\end{tablenotes}
\end{threeparttable}
\end{table*}

Table~\ref{tab:dsrg-mrpt2-ST-ccpvdz} reports adiabatic singlet-triplet splittings of the ten naphthyne isomers computed with the DSRG-MRPT2 approach using various basis sets (cc-pV$X$Z, with $X$ = D, T, Q, 5).
These results were computed using two active spaces:
1) CAS(2,2) which consists of two carbon $\sigma$ orbitals on radical centers and 2) CAS(12,12), which augments the CAS(2,2) space with ten carbon $\pi$ orbitals.

Following the analysis of Squires and Cramer,\cite{squires1998electronic} we separate the naphthyne isomers into three different groups characterized by different magnitudes of the singlet-triplet splitting.\cite{li2009energetics}
Group I naphthynes, which consists of (1,2) and (2,3)-naphthyne, have adjacent radical centers and their $\Delta_{\rm ST}$ is comparable to that of \textit{o}-benzyne ($-$37.5 $\pm$ 0.3 \kcal, from experiment).\cite{wenthold1998ultraviolet}
For group I naphthynes, through-bond interactions\cite{hoffmann1968benzynes, squires1998electronic} tend to stabilize the singlet state and are thus responsible for the relatively large $\Delta_{\rm ST}$ value.
Our best DSRG-MRPT2 estimates for the $\Delta E_{\rm ST}$ of (1,2) and (2,3)-naphthyne are $-$33.6 and $-$28.9 \kcal, respectively.

Group II contains (1,3)-naphthyne, the only isomer with the two radical centers in \textit{meta} position.  Our best estimate for the $\Delta_{\rm ST}$ value of this isomer is $-$14.2 \kcal, which is comparable to the value for \textit{m}-benzyne ($-$21.0 \kcal).  
Going from group I to II, there is a buildup of diradical character, which is reflected in the ratio between the two dominant configurations of the CAS(2,2) reference.
This quantity is reported at the bottom of Table~\ref{tab:dsrg-mrpt2-ST-ccpvdz}, and it goes from 2.6--2.5 for group I to 1.6 for group II naphthynes.
Group III naphthynes have singlet-triplet splittings that range from $-$1.0 to +4.1 \kcal.  This range is comparable to the $\Delta_{\rm ST}$ of \textit{p}-benzyne ($-$3.8 \kcal).
As indicated by the small $c_1/ c_2$ ratio, these species are almost pure diradicals.

\begin{table*}[ht!]
\begin{threeparttable}
\centering
\ifpreprint
\renewcommand{\arraystretch}{0.85}
\else
\renewcommand{\arraystretch}{1.2}
\fi
\caption{
Adiabatic singlet-triplet splittings ($\Delta E_{\rm ST} = E_{\rm S} - E_{\rm T}$) of naphthyne diradicals computed with the DSRG-MRPT2, CASPT2, and RMR-CCSD(T) approaches.
All DSRG-MRPT2 computations used triplet ROHF orbitals and the cc-pVDZ-RI auxiliary basis set.
RMR-CCSD(T) results used RHF and ROHF orbitals for singlet and triplet states, respectively.
CASPT2 results used CASSCF(12,12) orbitals.
RMR-CCSD(T) and DSRG-MRPT2 results are based on the same geometries (from DFT and CASSCF, see Sec.~\ref{sec:computational_details}), while CASPT2 results are based on CASSCF(12,12) optimized geometries.
The VDZ* basis set is constructed from the cc-pVDZ basis set by removing hydrogen $p$ functions.
} 
\label{tab:comparison-STccpvdz_star}
\ifpreprint
\begin{tabular*}{\columnwidth}{@{\extracolsep{\fill}}cc*{3}{d{3.1}}*{7}{d{2.1}}@{}}
\else
\begin{tabular*}{2\columnwidth}{@{\extracolsep{\fill}}cc*{3}{d{3.1}}*{7}{d{2.1}}@{}}
\fi
\hline
\hline
 & & \multicolumn{10}{c}{Naphthyne Isomers} \\
 \cline{3-12}
 & & \multicolumn{2}{c}{Group I} & \multicolumn{1}{c}{Group II} & \multicolumn{7}{c}{Group III} \\
  \cline{3-4} \cline{5-5} \cline{6-12} \vspace{-12pt} \\
CAS/Basis & Method & \multicolumn{1}{c}{1,2} & \multicolumn{1}{c}{2,3} & \multicolumn{1}{c}{1,3} & \multicolumn{1}{c}{1,5} & \multicolumn{1}{c}{1,6} & \multicolumn{1}{c}{1,4} & \multicolumn{1}{c}{2,7} & \multicolumn{1}{c}{2,6} & \multicolumn{1}{c}{1,7} & \multicolumn{1}{c}{1,8} \\
\hline
\multirow{2}{*}{(2,2)/VDZ*} & 
   RMR-CCSD(T)\tnote{a} & -35.2 & -28.9 & -12.7 & -0.3 & 2.1 & 1.6 & 2.9 & 5.7 & 6.5 & 6.5  \\
& DSRG-MRPT2\tnote{b} & -30.2 & -25.0 & -11.6 & 1.3 & 1.5  & 6.4 & 0.8 &   1.5 & 5.2 & 3.8 \\
\hline
\multirow{2}{*}{(12,12)/cc-pVDZ} & 
   CASPT2\tnote{c} & -31.8 & -27.7 & -17.5 & -8.6 & -2.2 & -6.7 & -3.9 & -3.0 & -2.7 & -1.8 \\
& DSRG-MRPT2\tnote{b} & -29.0 & -24.3 &-11.0 &   0.8& 2.0  & 5.3 &1.3& 1.3  & 5.2  & 4.3\\
\hline
\hline
\end{tabular*}
\begin{tablenotes}
\item [a] From Ref.~\citenum{li2009energetics}. 
\item [b] This work.
\item [c] From Ref.~\citenum{squires1998electronic}.  
\end{tablenotes}
\end{threeparttable}
\end{table*}

Table~\ref{tab:comparison-STccpvdz_star} reports a comparison between our CAS(2,2) DSRG-MRPT2 results and the reduced multireference coupled cluster with singles, doubles, and perturbative triples [RMR-CCSD(T)] results of Li and Paldus\cite{li2009energetics} using the same geometries and basis set.
Our DSRG-MRPT2 results for group I and II isomers agree very well with those from RMR-CCSD(T):
the maximum deviations are respectively 3.8 and 1.1 \kcal.
In the case of group III naphthynes, the disagreement between the DSRG-MRPT2 and RMR-CCSD(T) results is slightly less favorable.
The assignment of the ground state is consistent among the two methods, except for the (1,5) isomer, and (1,4)-naphthyne displays the largest absolute error (4.8 \kcal).

Table~\ref{tab:comparison-STccpvdz_star} also reports a comparison between our DSRG-MRPT2 results amd the CASPT2 results of Squires and Cramer,\cite{squires1998electronic} both obtained using a CAS(12,12) reference.
Note, that the comparison of these two sets of computations is complicated by the fact that the naphthynes geometries and orbitals used in these studies are different: the CASPT2 calculations use CASSCF(12,12) optimized geometries and orbitals.
As a consequence, the DSRG-MRPT2 results show some significant disagreements with the CASPT2 results.
For example, the DSRG-MRPT2 results favor triplet ground states for all the group III isomers, while CASPT2 predicts exactly the opposite.
Notice that the RMR-CCSD(T) approach also predicts triplet ground states for all group III naphthynes, except for the (1,5) isomer.

To illustrate the importance of the geometry used to compute $\Delta_{\rm ST}$, we optimized the singlet and triplet state geometry of (1,4)-naphthyne with the Mukherjee multireference coupled cluster approach with singles and doubles (Mk-MRCCSD) using the cc-pVDZ basis set and a CASSCF(2,2) reference.
At this level of theory, the singlet state is predicted to be the ground state and the adiabatic $\Delta_{\rm ST}$ = $-$4.98 \kcal.
The Mk-MRCCSD $\Delta_{\rm ST}$ of (1,4)-naphthyne agrees well with the experimental $\Delta_{\rm ST}$ of \textit{p}-benzyne, indicating that the nature of these two diradicals is similar.  More importantly, this result is also in agreement with the ground state assignment of CASPT2 computations.\cite{squires1998electronic}
\methodname $\Delta_{\rm ST}$ computed using the Mk-MRCCSD/cc-pVDZ geometries also favor a singlet ground state.
For example, when using ROHF orbitals, the \methodname $\Delta_{\rm ST}$ is equal to $-$0.9 \kcal [CAS(2,2)] and $-$1.98 \kcal [CAS(12,12)].
The use of CASSCF orbitals improves the agreement with the Mk-MRCCSD data: the corresponding $\Delta_{\rm ST}$ are $-$2.27 \kcal [CAS(2,2)] and $-$3.82 \kcal [CAS(12,12)].
As anticipated, ROHF orbitals tend to favor the triplet state, shifting $\Delta_{\rm ST}$ by $\sim$1.5 \kcal.
Although these results are not conclusive, they do suggest that to obtain reliable estimates of $\Delta_{\rm ST}$ for the naphthynes it is necessary to employ geometries optimized at a high level of theory.

\begin{table}[b]
\begin{threeparttable}
\centering
\ifpreprint
\renewcommand{\arraystretch}{0.85}
\else
\renewcommand{\arraystretch}{1.2}
\fi
\caption{A comparison of the vertical singlet-triplet splitting between MRCC and the DSRG-MRPT2.  All of these results use singlet geometries, RHF
orbitals, and a CAS(2,2).  }
\label{tab:mrccST}
\begin{tabular*}{\columnwidth}{@{\extracolsep{\stretch{1}}}ccd{2.2}d{1.2}*{2}{d{2.2}}@{}}
\hline
\hline
 &  & \multicolumn{4}{c}{Naphthyne Isomers} \\
 \cline{3-6}
Method & Basis &  \multicolumn{1}{c}{2,7} & \multicolumn{1}{c}{2,6} & \multicolumn{1}{c}{1,7} & \multicolumn{1}{c}{1,8} \\
\hline
\multirow{2}{*}{DSRG-MRPT2\tnote{a}}   
    & cc-pVDZ &  -3.0 & 0.2  & -1.6 & -2.6 \\
    & cc-pVTZ &  -3.2 & 0.2  & -1.7 & -2.8 \\
\hline
\multirow{2}{*}{BW-MRCCSD\tnote{b}} 
    & cc-pVDZ & -0.46 & 1.39 & -0.34 & -1.40 \\
    & cc-pVTZ & -0.78 & 1.05 & -0.50 & -1.47 \\
\hline
\multirow{2}{*}{Mk-MRCCSD\tnote{b}}
    & cc-pVDZ & 6.47 & 8.48 &7.29&  3.82 \\
    & cc-pVTZ & 7.16 & 9.79 &6.79&  4.43 \\
\hline
\hline
\end{tabular*}
\begin{tablenotes}
\item [a] This work.  
\item [b] From Ref.~\citenum{brabec2012towards}
\end{tablenotes}
\end{threeparttable}
\end{table}

In addition to adiabatic singlet-triplet splittings, in Table~\ref{tab:mrccST} we report a comparison of the vertical \methodname splittings with those from highly-accurate multireference coupled cluster (MRCC) computations by Brabec and coworkers.\cite{brabec2011massively}
These authors reported $\Delta_{\rm ST}$ for (2,7)-, (2,6)-, (1,7)- and (1,8)-naphthyne computed with the Brillouin--Wigner (BW) MRCC approach with
the \textit{a posteriori} correction\cite{Pittner1999p10275, hubavc2000size} and the Mk-MRCCSD approach.
To facilitate this comparison, all DSRG-MRPT2 results in Table~\ref{tab:mrccST} are computed using the same type of orbitals (restricted
Hartree--Fock) and geometries used by Brabec et al.\cite{brabec2011massively}
The DSRG-MRPT2 results agree well with those from BW-MRCCSD: both methods agree in the assignment of the ground state and the maximum error is only
2.54 \kcal.
Note, that there is a substantial disagreement between the BW- and Mk-MRCCSD results, which was attributed to the \textit{a posteriori} corrections used in BW-MRCCSD.

\subsection{Scaling with respect to basis set and active space size}

In this section we illustrate the efficiency of our DSRG-MRPT2 by reporting timings for the single-point energy computation of singlet (2,3)-naphthyne.
DSRG-MRPT2 timings for basis sets that range from 152 to 1240 orbitals are reported in Table~\ref{tab:basis_scaling}.
Due to the efficiency of the DF approximation, DSRG-MRPT2 computations with 1000--1500 may be performed routinely.
Indeed, our largest calculation using a CAS(2,2) reference and the cc-pV5Z basis set takes about 5 minutes with 8 threads on an Intel Xeon E5-2650 v2 processor.
This time is only about 5\% of the total time required (110 minutes), with the majority of the remaining part of the computation spent building the
Fock matrix (20 minutes) and the generation of the MO transformed DF integrals (35 minutes).  
The timings for the CAS(2,2) computations as function of the basis set size nicely follow the quadratic scaling expected from the DSRG-MRPT2 equations when the number of core and active orbitals is kept fixed.
Going from the CAS(2,2) to the CAS(12,12) active space we notice an increase of a factor 3--4 of the timing for the DSRG-MRPT2 step.
This result is significant because it suggests that for the active space here considered, terms that scale as a power of the number of active orbitals have a very small prefactor.
Indeed, even with the CAS(12,12) reference, the most expensive steps in the energy computation are the generation of the amplitudes 
$\tens{t}{ij}{ab}$ and $\tens{\tilde{v}}{ef}{mn}$, which require respectively $52 \%$ and $27 \%$ of the total time. 

\begin{table}[]
\begin{threeparttable}
\centering
\ifpreprint
\renewcommand{\arraystretch}{0.85}
\else
\renewcommand{\arraystretch}{1.2}
\fi
\caption{Timing of DSRG-MRPT2 naphthyne computations ($T_{\rm PT2}$, in seconds) as a function of basis set size ($N$). The total time ($T$) includes the CASCI step, generation of the DF integrals, and evaluation of the DSRG-MRPT2 energy.
These computations ran on one Intel Xeon E5-2650 v2 processor using 8 threads.}
\label{tab:basis_scaling}
\begin{tabular*}{\columnwidth}{@{\extracolsep{\stretch{1}}}*{2}{c}rd{3.1}d{2.1}@{}}
\hline
\hline
Active Space  & Basis & $N$ & \multicolumn{1}{c}{$T_{\rm PT2}$} & \multicolumn{1}{c}{$T_{\rm PT2} / T$ \%} \\
\hline
\multirow{4}{*}{CAS(2,2)}
& cc-pVDZ  & 170  & 3.5   & 21.0\\
& cc-pVTZ  & 384  & 22.6  & 18.6  \\
& cc-pVQZ  & 730  & 93.6  & 7.8\\
& cc-pV5Z  & 1240 & 316.5 & 4.8 \\
\hline
\multirow{3}{*}{CAS(12,12)}
& cc-pVDZ  & 170  & 13.2   & 9.6 \\
& cc-pVTZ  & 384  & 72.1   & 27.8\\
& cc-pVQZ  & 730  & 284.3  & 26.5\\
\hline
\hline
\end{tabular*}
\end{threeparttable}
\end{table}

\section{Conclusion}

In this work, we presented a new formulation of the DSRG-MRPT2 approach that takes advantage of two-electron integral factorization and the structure of CAS density matrices.
The resulting algorithm is similar to the one used in the evaluation of the single-reference MP2 energy, has reduced memory requirements, and allows the routine application of the DSRG-MRPT2 to systems with up to 50 atoms (1500--2000 basis functions).

To demonstrate the applicability of this novel DSRG-MRPT2 implementation to medium-sized system we studied the singlet-triplet splittings for the ten isomers of naphthyne diradicals. 
We reported computations with CAS(2,2) and CAS(12,12) active spaces and up to quintuple-$\zeta$ quality basis sets (1240 basis functions).
Overall, the DSRG-MRPT2 results are in good agreement with previously reported adiabatic singlet-triplet splittings computed at the RMR-CCSD(T)/VDZ* level of theory: the mean absolute deviation between the two approaches in only 2.7 \kcal.
We find that the singlet-triplet splittings of Group III naphthynes are strongly dependent on the quality of the molecular geometries.
This fact makes the comparison with previously reported CASPT2 results more difficult to analyze.
It also suggests that extra caution is required to interpret highly-correlated results for naphthynes based on DFT or CASSCF geometries.
DSRG-MRPT2 computations with larger bases suggest that one should at least use a triple-$\zeta$ basis set to converge the single-triplet splitting of Group III naphthynes to 0.4 \kcal, while a quadruple-$\zeta$ basis is necessary to reduce this error to about 0.1 \kcal.

In this work we have showed that the cost of evaluating the DSRG-MRPT2 energy can be significantly reduced by resorting to integral factorization techniques.
Nevertheless, the computational scaling of integral-factorized \methodname remains proportional to the fifth power of the number of electrons.
Therefore, to apply this approach to systems with 100--150 atoms it will be necessary to reduce its computational scaling.
Given the simplicity of the \methodname equations, an interesting option is to combine the prescreening of atomic orbital (AO) integrals with Laplace transformation of the energy denominators.\cite{Almlof:1991ki,Haser:1992bh,ayala1999linear,Lambrecht:2005df,doser2009linear}
One novel issue that arises in the application of the Laplace transformation to the \methodname approach is the fact that the energy denominators are renormalized.
However, we think that this problem may be addressed either by finding a suitable decomposition of the renormalized denominators, or by redefining the source operator to treat the most expensive contributions (from diagram F1--F3) as non-renormalized quantities.
We anticipate that the Laplace-transformed AO-DSRG-MRPT2 will be an essential tool to go beyond the current limit of 2000 basis functions.

\begin{acknowledgments}

The authors are grateful to Dr.\ Robert M.\ Parrish for many insightful discussions. 
K.~P.~H.~would like to thank the entire Evangelista lab for their insightful advice.  
This research was supported by start-up funds provided by Emory University.

\end{acknowledgments}

\bibliography{DFDSRG}
\end{document}